\documentstyle[12pt]{article}
\begin{document}
\begin{center}
{\bf \large Scaling Relation for Leptonic Constants\\ of Higher Excitations in
Heavy Quarkonium}
\vspace*{1cm}
V.V.Kiselev\\
{\it Institute for High Energy Physics,\\
Protvino, Moscow Region, 142284, Russia,\\
E-mail: kiselev@mx.ihep.su\\
Fax: +7-095-230-23-37}
\end{center}
\begin{abstract}

\end{abstract}

\section*{Introduction}

A description of quark bound states demands an application of
nonperturbative approaches in QCD, which comes in the strong
coupling regime ($\alpha_S \sim 1$) within the region of the hadronization
at large distances ($r \sim 1/\Lambda_{QCD}$, $\Lambda_{QCD} \sim 200$ MeV).

Considering the bound states with the heavy quarks ($m_Q \gg \Lambda_{QCD}$),
one derives some regularities, simplifying the solution of problem.

For the hadrons with a single heavy quark, the virtualities of the heavy
quark are low ($\sim \Lambda_{QCD}$) and one can make the expansion of the
heavy quark QCD action over the small parameter $\Lambda_{QCD}/m_Q$.
In the leading approximation, the effective action possesses the symmetry
with respect to the substitution of a heavy quark, moving with a
velocity $\vec{v}$, by any other heavy quark, moving with the same
velocity $\vec{v}$ and having an arbitrary orientation of its spin \cite{1}.
The symmetry allows one to state the scaling law for leptonic constants of
the heavy mesons, containing a single heavy quark, and the universal
dependence of the form factors for exclusive semileptonic weak transitions
between the heavy hadrons such as $B \to D^{(*)}l \nu$, so that the universal
function has fixed normalization at the zero recoil point.

For the heavy quarkonioum $(Q\bar Q')$, the nonrelativistic heavy quark motion
inside the bound state has allowed one to develop the approach of
phenomenological potential models. The QCD-motivated potentials combine the
linear rise of the quark interaction energy at the large distances and
the Coulomb-like interaction at the small distances \cite{2}.
Such models reach the high accuracy in the description of the
mass spectra of heavy quarkonia (the ($c\bar c$) charmonium and the
($b\bar b$) bottomonium), $\delta m \sim 30$ MeV, however, their accuracy
in the description of leptonic constants is very low ($\sim 25 \%$).

Another powerful tool for the description of bound states with
the heavy quarks has become the QCD sum rules \cite{3}, combining
perturbative calculations and an account of contributions by the
vacuum expectation values of composite operators, i.e. by the
quark gluon condensates such as $<m \bar q q>$, $<\alpha_s G^2_{\mu\nu}>$ and
so on. However, making the consideration with the finite number of
terms in the QCD perturbation theory for the Wilson's coefficients and
taking into the account only the restricted set of quark-gluon condensates,
results of the QCD sum rules get unphysical dependence on an external
parameter of the sum rule scheme (the number of spectral density
moments $n$ or the Borel transformation parameter $\sigma$). This dependence
essentially decreases the predictive power of QCD sum rules. An
additional parameter is also the threshold $s_{th}$, discriminating the
resonant region from the hadronic continuum. Moreover, the weight functions,
rapidly dropping with the energy increase, define the averaging scheme of
QCD sum rules and do not allow one to draw some conclusions on the
contributions by the higher excitations of quarkonium with the given
quantum numbers, so that  these contributions are practically neglected.

The QCD sum rule scheme, using the data on the quarkonium mass spectrum
has been recently offered in ref.\cite{4}. In the scheme, the conditions
of nonrelativistic quark motion and the small value of ratio
$\Lambda_{QCD}/m_Q$ have allowed one to state the scaling relation
for the leptonic constants of $S$-wave quarkonia ($Q\bar Q$)
\begin{equation}
\frac{f^2}{M} = const.\;, \label{h1}
\end{equation}
independently of the heavy quark flavours. The generalization of eq.(\ref{h1})
for the heavy quarkonium ($Q\bar Q'$) has been considered in ref.\cite{5},
so that
\begin{equation}
\frac{f^2}{M}\;\biggl(\frac{M}{4\mu}\biggr)^2 = const.\;, \label{2}
\end{equation}
where $\mu=m_Q m_{Q'}/(m_Q+m_{Q'})$ is the reduced mass of ($Q\bar Q'$)
system.

In the present paper we use the QCD sum rule scheme of refs.\cite{4,5}
to derive the relation for the leptonic constants of $nS$-states
of heavy quarkonium
\begin{equation}
\frac{f^2_{n_1}}{f^2_{n_2}} = \frac{n_2}{n_1}\;, \label{3}
\end{equation}
that does not depend on the heavy quarkonium content.

In Section 1 we describe the QCD sum rule scheme, using the spectroscopic data,
and derive eq.(\ref{3}). In Section 2 we make the phenomenological analysis
of relation (\ref{3}) and show, that it gives a good description of the
experimental relations for the leptonic constants in the $\psi$- and
$\Upsilon$-particle families.
In the Conclusion we discuss the obtained results.

\section{QCD Sum Rules for Heavy Quarkonium}

Let us consider the two-point correlator functions of quark currents
\begin{eqnarray}
\Pi_{\mu\nu} (q^2) & = & i \int d^4x e^{iqx} <0|T J_{\mu}(x)
J^{\dagger}_{\nu}(0)|0>\;,
\label{1} \\
\Pi_P (q^2) & = & i \int d^4x e^{iqx} <0|T J_5(x) J^{\dagger}_5(0)|0>\;,
\end{eqnarray}
where
\begin{eqnarray}
J_{\mu}(x) & = & \bar Q_1(x) \gamma_{\mu} Q_2(x)\;,\\
J_5(x) & = & \bar Q_1(x) \gamma_5 Q_2(x)\;,\\
\end{eqnarray}
$Q_i$ is the spinor field of the heavy quark with $i = c, b$.

Further, write down
\begin{equation}
\Pi_{\mu\nu} = \biggl(-g_{\mu\nu}+\frac{q_{\mu} q_{\nu}}{q^2}\biggr) \Pi_V(q^2)
+ \frac{q_{\mu} q_{\nu}}{q^2} \Pi_S(q^2)\;,
\end{equation}
where $\Pi_V$ and $\Pi_S$ are the vector and scalar correlator functions,
respectively. In what follows we will consider the vector and pseudoscalar
correlators: $\Pi_V(q^2)$ and $\Pi_P(q^2)$.

Define the leptonic constants $f_V$ and $f_P$
\begin{eqnarray}
<0|J_{\mu}(x) |V(\lambda)> & = & i \epsilon^{(\lambda)}_{\mu}\;
f_V M_V\;e^{ikx}\;,\\
<0|J_{5\mu}(x)|P> & = & i k_{\mu}\;f_P e^{ikx}\;,
\end{eqnarray}
where
\begin{equation}
J_{5\mu}(x)  =  \bar Q_1(x) \gamma_5 \gamma_{\mu} Q_2(x)\;,
\end{equation}
so that
\begin{equation}
<0|J_{5}(x)|P>  =  i\;\frac{f_P M_P^2}{m_1+m_2}\;e^{ikx}\;, \label{9}
\end{equation}
where $|V>$ and  $|P>$ are the state vectors of $1^-$ and $0^-$
quarkonia, and $\lambda$ is the vector quarkonium polarization, $k$
is 4-momentum of the meson, $k_{P,V}^2 = M_{P,V}^2$.

Considering the charmonium ($\psi$, $\psi '$ ...) and bottomonium ($\Upsilon$,
$\Upsilon '$, $\Upsilon ''$ ...), one can easily show that the relation
between the width of
leptonic decay $V \to e^+ e^-$  and $f_V$ has the form
\begin{equation}
\Gamma (V \to e^+ e^-) = \frac{4 \pi}{9}\;e_i^2 \alpha_{em}^2\;
\frac{f_V^2}{M_V}\;,
\end{equation}
where $e_i$ is the electric charge of quark $i$.

In the region of narrow nonoverlapping resonances, it follows from
eqs.(\ref{1}) - (\ref{9}) that
\begin{eqnarray}
\frac{1}{\pi} \Im m \Pi_V^{(res)} (q^2) & = &
\sum_n f_{Vn}^2 M_{Vn}^2 \delta(q^2-M_{Vn}^2)\;,
\label{11} \\
\frac{1}{\pi} \Im m \Pi_P^{(res)} (q^2) & = &
\sum_n f_{Pn}^2 M_{Pn}^4\;\frac{1}{(m_1+m_2)^2} \delta(q^2-M_{Pn}^2)\;.
\end{eqnarray}
Thus, for the observed spectral function one has
\begin{equation}
\frac{1}{\pi} \Im m \Pi_{V,P}^{(had)} (q^2)  = \frac{1}{\pi} \Im m
\Pi_{V,P}^{(res)} (q^2)+ \rho_{V,P}(q^2, \mu_{V,P}^2)\;,
\label{13}
\end{equation}
where $\rho (q^2,\;\mu^2)$ is the continuum contribution, which is
not equal to zero at $q^2 > \mu^2$.

Moreover, the operator product expansion gives
\begin{equation}
\Pi^{(QCD)} (q^2)  = \Pi^{(pert)} (q^2)+ C_G(q^2) <\frac{\alpha_S}{\pi} G^2> +
C_i(q^2)<m_i \bar Q_i Q_i>+ \dots\;,
\label{14}
\end{equation}
where the perturbative contribution $\Pi^{(pert)}(q^2)$ is labeled, and
the nonperturbative one is expressed in the form of sum
of quark-gluon condensates
with the Wilson's coefficients, which can be calculated in the QCD
perturbative theory.

In eq.(\ref{14}) we have been restricted by the contribution of vacuum
expectation values for the operators with dimension $d =4$.
For $C^{(P)}_G (q^2)$ one has, for instance, \cite{3}
\begin{equation}
C_G^{(P)} = \frac{1}{192 m_1 m_2}\;\frac{q^2}{\bar q^2}\;
\biggl(\frac{3(3v^2+1)(1-v^2)^2}
{2v^5} \ln \frac{1+v}{1-v} - \frac{9v^4+4v^2+3}{v^4}\biggr)\;, \label{15}
\end{equation}
where
\begin{equation}
\bar q^2 = q^2 - (m_1-m_2)^2\;,\;\;\;\;v^2 = 1-\frac{4m_1 m_2}{\bar q^2}\;.
\label{16}
\end{equation}
The analogous formulae for other Wilson's coefficients can be found in
Ref.\cite{3}. In what follows it will be clear that the explicit form
of coefficients has no significant meaning for the present consideration.

In the leading order of QCD perturbation theory it has been found for
the imaginary part of correlator that \cite{3}
\begin{eqnarray}
\Im m \Pi_V^{(pert)} (q^2) & = & \frac{\tilde s}{8 \pi s^2}
(3 \bar s s - \bar s^2 + 6m_1 m_2 s - 2 m_2^2 s) \theta(s-(m_1+m_2)^2),\\
\Im m \Pi_P^{(pert)} (q^2) & = & \frac{3 \tilde s}{8 \pi s^2}
(s - (m_1-m_2)^2) \theta(s-(m_1+m_2)^2)\;,
\end{eqnarray}
where $\bar s = s-m_1^2+m_2^2$, $ \tilde s^2 = \bar s^2 -4 m_2^2 s$.

The one-loop contribution into $\Im m \Pi(q^2)$ can be included into the
consideration (see, for example, Ref.\cite{3}). However, we note that the
more essential correction is that of summing a set over the powers of
$(\alpha_s/v)$, where $v$ is defined in eq.(\ref{16}) and is a relative quark
velocity, and $\alpha_S$ is the QCD interaction constant. In Ref.\cite{3}
it has been shown that account of the Coulomb-like gluonic
interaction between the quarks leads to the factor
\begin{equation}
F(v) = \frac{4 \pi}{3}\;\frac{\alpha_S}{v}\; \frac{1}{1-\exp (-\frac{4 \pi
\alpha_S}{3 v})}\;,
\end{equation}
so that the expansion of the $F(v)$ over $\alpha_S/v \ll 1$ restores,
precisely,
the one-loop $O(\frac{\alpha_S}{v})$ correction
\begin{equation}
F(v) \approx 1 - \frac{2 \pi}{3}\;\frac{\alpha_s}{v}\; \dots \label{20}
\end{equation}
In accordance with the dispersion relation one has the QCD sum rules,
which state that, in average, it is true that, at least, at $q^2 < 0$
\begin{equation}
\frac{1}{\pi}\;\int\frac{\Im m \Pi^{(had)}(s)}{s-q^2} ds = \Pi^{(QCD)}(q^2)\;,
 \label{21}
\end{equation}
where the necessary subtractions are omitted. $\Im m \Pi^{(had)}(q^2)$ and
 $\Pi^{(QCD)}(q^2)$ are defined by eqs.(\ref{11}) - (\ref{13}) and
eqs.(\ref{14}) - (\ref{20}), respectively.
eq.(\ref{21}) is the base to develop the sum rule approach in the forms
of the correlator function moments and of the Borel transform analysis
(see Ref.\cite{3}). The truncation of the set in the right hand side of
eq.(\ref{21}) leads to the mentioned unphysical dependence of the $f_{P,V}$
values on the external parameter of the sum rule scheme.

Further, let us use the conditions, simplifying the consideration due to
the heavy quarkonium.

\subsection{Nonperturbative Contribution}

We assume that, in the limit of the very heavy quark mass, the power
corrections of nonperturbative contribution are small. From eq.(\ref{15})
one can see that, for example,
\begin{equation}
C_G^{(P)}(q^2) \approx O(\frac{1}{m_1 m_2})\;,\;\; \Lambda/m_{1,2}\ll 1\;,
\end{equation}
where $v$ is fixed,  $q^2 \sim (m_1 + m_2)^2$,
when $\Im m \Pi^{(pert)}(q^2) \sim (m_1+m_2)^2$.
It is evident that, due to the purely dimensional consideration, one can
believe that the Wilson's coefficients tend to zero as
$1/m_{1,2}^2$.

Thus, the limit of very large heavy quark mass implies that one can neglect
the quark-gluon condensate contribution.

\subsection{Nonrelativistic Quark Motion}

The nonrelativistic quark motion implies that, in the resonant region, one has,
in accordance with eq.(\ref{16}),
\begin{equation}
v \to 0\;.
\end{equation}
So, one can easily find that in the leading order
\begin{equation}
\Im m \Pi_P^{(pert)}(s) \approx  \Im m \Pi_V^{(pert)}(s) \to \frac {3 v}
{8 \pi^2} s\; \biggl(\frac{4\mu}{M}\biggr)^2\;,
\end{equation}
so that with account of the Coulomb factor
\begin{equation}
F(v) \simeq \frac{4 \pi}{3}\; \frac{\alpha_S}{v}\;,
\end{equation}
one obtaines
\begin{equation}
\Im m \Pi_{P,V}^{(pert)}(s) \simeq \frac{\alpha_S}{2} s\;
\biggl(\frac{4\mu}{M}\biggr)^2\;. \label{27}
\end{equation}

\subsection{"Smooth Average Value" Scheme of the Sum Rules}

As for the hadronic part of the correlator, one can write down for the narrow
resonance contribution
\begin{eqnarray}
\Pi_V^{(res)}(q^2) & = & \int \frac{ds}{s-q^2}\;\sum_n f^2_{Vn} M^2_{Vn}
\delta(s-M_{Vn}^2)\;,
\label{28} \\
\Pi_P^{(res)}(q^2) & = & \int \frac{ds}{s-q^2}\;\sum_n f^2_{Pn}
\frac{M^4_{Pn}}{(m_1+m_2)^2} \delta(s-M_{Pn}^2)\;,\label{29}
\end{eqnarray}
The integrals in eqs.(\ref{28})-(\ref{29}) are simply calculated, and
this procedure is generally used.

In the presented scheme, let us introduce the function of state number
$n(s)$, so that
\begin{equation}
n(m_k^2) = k\;.
\end{equation}
This definition seems to be reasonable in the resonant region.
Then one has, for example, that
\begin{equation}
\frac{1}{\pi}\; \Im m \Pi_V^{(res)}(s) = s f^2_{Vn(s)}\; \frac{d}{ds} \sum_k
\theta(s-M^2_{Vk})\;.
\end{equation}
Further, it is evident that
\begin{equation}
\frac{d}{ds} \sum_k \theta(s-M_k^2) = \frac{dn(s)}{ds}\;\frac{d}{dn} \sum_k
\theta(n-k)\;,
\end{equation}
and eq.(\ref{28}) can be rewritten as
\begin{equation}
\Pi_V^{(res)}(q^2) = \int \frac{ds}{s-q^2}\; s f^2_{Vn(s)}\;\frac{dn(s)}{ds}\;
\frac{d}{dn} \sum_k \theta(n-k)\;.
\end{equation}
The "smooth average value" scheme means that
\begin{equation}
\Pi_V^{(res)}(q^2) = <\frac{d}{dn} \sum_k \theta(n-k)>\; \int \frac{ds}{s-q^2}
s f^2_{Vn(s)} \frac{dn(s)}{ds}\;.
\end{equation}
It is evident that, in average, the first derivative of step-like function
in the resonant region is equal to
\begin{equation}
<\frac{d}{dn} \sum_k \theta(n-k)> \simeq 1\;.
\end{equation}
Thus, in the scheme one has
\begin{eqnarray}
<\Pi_V^{(res)}(q^2)> & \approx & \int \frac{ds}{s-q^2}
s f^2_{Vn(s)}\; \frac{dn(s)}{ds}\;,
\label{35} \\
<\Pi_P^{(res)}(q^2)> & \approx & \int \frac{ds}{s-q^2}
\frac{s^2 f^2_{Pn(s)}}{(m_1+m_2)^2}\; \frac{dn(s)}{ds}\;.
\label{36}
\end{eqnarray}
Eqs.(\ref{35})-(\ref{36}) give the average correlators for the vector and
pseudoscalar mesons, therefore, due to eq.(\ref{21}) we state  that
\begin{equation}
\Im m <\Pi^{(hadr)}(q^2)> = \Im m \Pi^{(QCD)}(q^2)\;,
\end{equation}
that gives with account of eqs.(\ref{27}), (\ref{35}) and
(\ref{36}) at the physical points $s_n =M_n^2$
\begin{equation}
\frac{f_n^2}{M_n} = \frac{\alpha_S}{\pi} \; \frac{dM_n}{dn}
\; \biggl(\frac{4\mu}{M}\biggr)^2\;, \label{38}
\end{equation}
where in the limit of heavy quarks we use, that for the resonances
one has
\begin{equation}
m_1 +m_2 \approx M\;,\label{39}
\end{equation}
so that
\begin{equation}
f_{Vn} \simeq f_{Pn} = f_n\;.\label{40}
\end{equation}
Thus, one can conclude that for the heavy quarkonia the QCD sum rules give
the identity of $f_P$ and $f_V$ values for the pseudoscalar
and vector states.

Eq.(\ref{38}) differs from the ordinary sum rule scheme because it does not
contain the parameters, which are external to QCD. The quantity
$dM_n/dn$ is purely phenomenological. It defines the average mass difference
between the nearest levels with the identical quantum numbers.

Further, as it has been shown in ref.\cite{6}, in the region of average
distances between the heavy quarks in the charmonium and the bottomonium,
\begin{equation}
0.1\; fm < r < 1\;fm\;, \label{2.1}
\end{equation}
the QCD-motivated potentials allow the approximation in the form of
logarithmic law \cite{7} with the simple scaling properties, so
\begin{equation}
\frac{dn}{dM_n} = const.\;,\label{2.2}
\end{equation}
i.e. the density of heavy quarkonium states with the given quantum
numbers do not depend on the heavy quark flavours.

In ref.\cite{5} it has been shown, that relation (\ref{2.2}) is also
practically valid for the heavy quark potential approximation by the
power law (Martin potential) \cite{8}, where, neglecting a low value of
the binding energy for the quarks inside the quarkonium, one can again
get eq.(\ref{2.2}).

In ref.\cite{4} it has been found, that relation (\ref{2.2}) is valid
 with the accuracy up to small logarithmic corrections over the
reduced mass of quarkonium, if one makes the quantization of
$S$-wave states for the quarkonium with the Martin potential by the
Bohr-Sommerfeld procedure.

Moreover, with the accuracy up to the logarithmic corrections, $\alpha_S$
is the constant value. Thus, as it has been shown in refs.\cite{4,5},
for the leptonic constants of $S$-wave quarkonia, the scaling relation
takes place
\begin{equation}
\frac{f^2}{M}\; \biggl(\frac{M}{4\mu}\biggr)^2 = const.\;, \label{2.3}
\end{equation}
independently of the heavy quark flavours.

Taking into the account eqs.(\ref{39}) and (\ref{40}) and integrating
eqs.(\ref{35}), (\ref{36}) by parts, one can get with the accuracy up to
border terms, that one has
\begin{equation}
-2f_n\; \frac{df_n}{dn}\; \frac{dn}{dM_n}\; n = \frac{\alpha_s}{\pi}\;
M_n\; \biggl(\frac{4\mu}{M_n}\biggr)^2\;. \label{2.4}
\end{equation}
Comparing eqs.(\ref{38}) and (\ref{2.4}), one finds
\begin{equation}
\frac{df_n}{f_n dn} = - \frac{1}{2n}\;, \label{2.5}
\end{equation}
that  gives, after the integration, eq.(\ref{3}):
$$\frac{f^2_{n_1}}{f^2_{n_2}} = \frac{n_2}{n_1}\;.
$$
Relation (\ref{3}) leads to that the border terms, which have been neglected
in the writing of eq.(\ref{2.4}), are identically equal to zero.

Thus, under the conditions of small $\Lambda_{QCD}/m_Q$ ratio value and
the nonrelativistic heavy quark motion, the described QCD sum rule scheme
with the "smooth average value" allows one to derive the scaling expression,
relating the leptonic constants of different $nS$-wave levels,
independently of the quark content of heavy quarkonium.

\section{Analysis of Scaling Relation}
\begin{table}[t]
\caption{The experimental values of leptonic constants (in MeV)
for the $nS$-bottomonia in comparison with the
estimates of present model.}
\label{th1}
\begin{center}
\begin{tabular}{||c|c|c||}
\hline
quantity & exp. & present \\
\hline
$f_1$ & $715\pm15$ & input \\
$f_2$ & $487\pm16$ & $506\pm10$ \\
$f_3$ & $429\pm14$ & $412\pm8$ \\
$f_4$ & $320\pm30$ & $358\pm7$ \\
$f_5$ & $369\pm46$ & $320\pm7$ \\
$f_6$ & $240\pm30$ & $292\pm6$ \\
\hline
\end{tabular}
\end{center}
\end{table}

First, note that eq.(\ref{2.3}), relating the leptonic constants of
different quarkonia, turns out to be certainly valid
for the quarkonia with the hidden
flavour ($c\bar c$, $b\bar b$), where $4\mu/M=1$ (see \cite{4}).
In that case, to estimate
the constant value in the right hand side of eq.(\ref{2.3}) we have
supposed in refs.\cite{4,5}, that $\alpha_S$ has the value, defining the
Coulomb part of potential in the Cornell model (see \cite{2}), and
the $<dM/dn>$ value is equal to the average distance between the nearest
$S$-wave levels in the bottomonium.

\begin{table}[b]
\caption{The experimental values of leptonic constants (in MeV)
for the $nS$-charmonia in comparison with the
estimates of present model.}
\label{th2}
\begin{center}
\begin{tabular}{||c|c|c||}
\hline
quantity & exp. & present \\
\hline
$f_1$ & $410\pm14$ & input \\
$f_2$ & $283\pm14$ & $290\pm10$ \\
$f_3$ & $205\pm20$ & $237\pm8$ \\
$f_4$ & $180\pm30$ & $205\pm7$ \\
$f_5$ & $145\pm15$ & $183\pm6$ \\
\hline
\end{tabular}
\end{center}
\end{table}
\setlength{\unitlength}{0.85mm}\thicklines
\begin{figure}[t]
\begin{center}
\begin{picture}(100,95)
\put(15,10){\framebox(80,80)}
\put(15,30){\line(1,0){3}}
\put(3,30){$200$}
\put(15,50){\line(1,0){3}}
\put(3,50){$400$}
\put(15,70){\line(1,0){3}}
\put(3,70){$600$}
\put(15,20){\line(1,0){3}}
\put(15,40){\line(1,0){3}}
\put(15,60){\line(1,0){3}}
\put(15,80){\line(1,0){3}}

\put(0,93){$f_n,\;MeV$}

\put(25,10){\line(0,1){3}}
\put(35,10){\line(0,1){3}}
\put(45,10){\line(0,1){3}}
\put(55,10){\line(0,1){3}}
\put(65,10){\line(0,1){3}}
\put(75,10){\line(0,1){3}}
\put(85,10){\line(0,1){3}}
\put(25,2){$1$}
\put(35,2){$2$}
\put(45,2){$3$}
\put(55,2){$4$}
\put(65,2){$5$}
\put(75,2){$6$}
\put(85,2){$7$}
\put(95,2){$n$}

\put(25,81.5){\circle*{1.6}}
\put(35,58.7){\circle*{1.6}}
\put(45,52.9){\circle*{1.6}}
\put(55,42.0){\circle*{1.6}}
\put(65,46.9){\circle*{1.6}}
\put(75,34.0){\circle*{1.6}}
\put(55,39.0){\line(0,1){6}}
\put(65,42.3){\line(0,1){9.2}}
\put(75,31.0){\line(0,1){6}}

 \put( 25.10, 81.15){\circle*{0.75}}
 \put( 25.20, 80.80){\circle*{0.75}}
 \put( 25.30, 80.45){\circle*{0.75}}
 \put( 25.40, 80.11){\circle*{0.75}}
 \put( 25.50, 79.78){\circle*{0.75}}
 \put( 25.60, 79.45){\circle*{0.75}}
 \put( 25.70, 79.12){\circle*{0.75}}
 \put( 25.80, 78.80){\circle*{0.75}}
 \put( 25.90, 78.48){\circle*{0.75}}
 \put( 26.00, 78.17){\circle*{0.75}}
 \put( 26.10, 77.86){\circle*{0.75}}
 \put( 26.20, 77.56){\circle*{0.75}}
 \put( 26.30, 77.26){\circle*{0.75}}
 \put( 26.40, 76.97){\circle*{0.75}}
 \put( 26.50, 76.67){\circle*{0.75}}
 \put( 26.60, 76.39){\circle*{0.75}}
 \put( 26.70, 76.10){\circle*{0.75}}
 \put( 26.80, 75.82){\circle*{0.75}}
 \put( 26.90, 75.54){\circle*{0.75}}
 \put( 27.00, 75.27){\circle*{0.75}}
 \put( 27.10, 75.00){\circle*{0.75}}
 \put( 27.20, 74.73){\circle*{0.75}}
 \put( 27.30, 74.47){\circle*{0.75}}
 \put( 27.40, 74.21){\circle*{0.75}}
 \put( 27.50, 73.95){\circle*{0.75}}
 \put( 27.60, 73.70){\circle*{0.75}}
 \put( 27.70, 73.45){\circle*{0.75}}
 \put( 27.80, 73.20){\circle*{0.75}}
 \put( 27.90, 72.95){\circle*{0.75}}
 \put( 28.00, 72.71){\circle*{0.75}}
 \put( 28.10, 72.47){\circle*{0.75}}
 \put( 28.20, 72.23){\circle*{0.75}}
 \put( 28.30, 72.00){\circle*{0.75}}
 \put( 28.40, 71.77){\circle*{0.75}}
 \put( 28.50, 71.54){\circle*{0.75}}
 \put( 28.60, 71.31){\circle*{0.75}}
 \put( 28.70, 71.09){\circle*{0.75}}
 \put( 28.80, 70.86){\circle*{0.75}}
 \put( 28.90, 70.65){\circle*{0.75}}
 \put( 29.00, 70.43){\circle*{0.75}}
 \put( 29.10, 70.21){\circle*{0.75}}
 \put( 29.20, 70.00){\circle*{0.75}}
 \put( 29.30, 69.79){\circle*{0.75}}
 \put( 29.40, 69.58){\circle*{0.75}}
 \put( 29.50, 69.38){\circle*{0.75}}
 \put( 29.60, 69.17){\circle*{0.75}}
 \put( 29.70, 68.97){\circle*{0.75}}
 \put( 29.80, 68.77){\circle*{0.75}}
 \put( 29.90, 68.58){\circle*{0.75}}
 \put( 30.00, 68.38){\circle*{0.75}}
 \put( 30.10, 68.19){\circle*{0.75}}
 \put( 30.20, 67.99){\circle*{0.75}}
 \put( 30.30, 67.80){\circle*{0.75}}
 \put( 30.40, 67.62){\circle*{0.75}}
 \put( 30.50, 67.43){\circle*{0.75}}
 \put( 30.60, 67.25){\circle*{0.75}}
 \put( 30.70, 67.06){\circle*{0.75}}
 \put( 30.80, 66.88){\circle*{0.75}}
 \put( 30.90, 66.70){\circle*{0.75}}
 \put( 31.00, 66.53){\circle*{0.75}}
 \put( 31.10, 66.35){\circle*{0.75}}
 \put( 31.20, 66.18){\circle*{0.75}}
 \put( 31.30, 66.00){\circle*{0.75}}
 \put( 31.40, 65.83){\circle*{0.75}}
 \put( 31.50, 65.66){\circle*{0.75}}
 \put( 31.60, 65.49){\circle*{0.75}}
 \put( 31.70, 65.33){\circle*{0.75}}
 \put( 31.80, 65.16){\circle*{0.75}}
 \put( 31.90, 65.00){\circle*{0.75}}
 \put( 32.00, 64.84){\circle*{0.75}}
 \put( 32.10, 64.68){\circle*{0.75}}
 \put( 32.20, 64.52){\circle*{0.75}}
 \put( 32.30, 64.36){\circle*{0.75}}
 \put( 32.40, 64.20){\circle*{0.75}}
 \put( 32.50, 64.05){\circle*{0.75}}
 \put( 32.60, 63.90){\circle*{0.75}}
 \put( 32.70, 63.74){\circle*{0.75}}
 \put( 32.80, 63.59){\circle*{0.75}}
 \put( 32.90, 63.44){\circle*{0.75}}
 \put( 33.00, 63.29){\circle*{0.75}}
 \put( 33.10, 63.15){\circle*{0.75}}
 \put( 33.20, 63.00){\circle*{0.75}}
 \put( 33.30, 62.85){\circle*{0.75}}
 \put( 33.40, 62.71){\circle*{0.75}}
 \put( 33.50, 62.57){\circle*{0.75}}
 \put( 33.60, 62.43){\circle*{0.75}}
 \put( 33.70, 62.29){\circle*{0.75}}
 \put( 33.80, 62.15){\circle*{0.75}}
 \put( 33.90, 62.01){\circle*{0.75}}
 \put( 34.00, 61.87){\circle*{0.75}}
 \put( 34.10, 61.74){\circle*{0.75}}
 \put( 34.20, 61.60){\circle*{0.75}}
 \put( 34.30, 61.47){\circle*{0.75}}
 \put( 34.40, 61.33){\circle*{0.75}}
 \put( 34.50, 61.20){\circle*{0.75}}
 \put( 34.60, 61.07){\circle*{0.75}}
 \put( 34.70, 60.94){\circle*{0.75}}
 \put( 34.80, 60.81){\circle*{0.75}}
 \put( 34.90, 60.69){\circle*{0.75}}
 \put( 35.00, 60.56){\circle*{0.75}}
 \put( 35.10, 60.43){\circle*{0.75}}
 \put( 35.20, 60.31){\circle*{0.75}}
 \put( 35.30, 60.18){\circle*{0.75}}
 \put( 35.40, 60.06){\circle*{0.75}}
 \put( 35.50, 59.94){\circle*{0.75}}
 \put( 35.60, 59.82){\circle*{0.75}}
 \put( 35.70, 59.70){\circle*{0.75}}
 \put( 35.80, 59.58){\circle*{0.75}}
 \put( 35.90, 59.46){\circle*{0.75}}
 \put( 36.00, 59.34){\circle*{0.75}}
 \put( 36.10, 59.22){\circle*{0.75}}
 \put( 36.20, 59.11){\circle*{0.75}}
 \put( 36.30, 58.99){\circle*{0.75}}
 \put( 36.40, 58.88){\circle*{0.75}}
 \put( 36.50, 58.76){\circle*{0.75}}
 \put( 36.60, 58.65){\circle*{0.75}}
 \put( 36.70, 58.54){\circle*{0.75}}
 \put( 36.80, 58.43){\circle*{0.75}}
 \put( 36.90, 58.32){\circle*{0.75}}
 \put( 37.00, 58.21){\circle*{0.75}}
 \put( 37.10, 58.10){\circle*{0.75}}
 \put( 37.20, 57.99){\circle*{0.75}}
 \put( 37.30, 57.88){\circle*{0.75}}
 \put( 37.40, 57.77){\circle*{0.75}}
 \put( 37.50, 57.67){\circle*{0.75}}
 \put( 37.60, 57.56){\circle*{0.75}}
 \put( 37.70, 57.46){\circle*{0.75}}
 \put( 37.80, 57.35){\circle*{0.75}}
 \put( 37.90, 57.25){\circle*{0.75}}
 \put( 38.00, 57.15){\circle*{0.75}}
 \put( 38.10, 57.04){\circle*{0.75}}
 \put( 38.20, 56.94){\circle*{0.75}}
 \put( 38.30, 56.84){\circle*{0.75}}
 \put( 38.40, 56.74){\circle*{0.75}}
 \put( 38.50, 56.64){\circle*{0.75}}
 \put( 38.60, 56.54){\circle*{0.75}}
 \put( 38.70, 56.44){\circle*{0.75}}
 \put( 38.80, 56.35){\circle*{0.75}}
 \put( 38.90, 56.25){\circle*{0.75}}
 \put( 39.00, 56.15){\circle*{0.75}}
 \put( 39.10, 56.06){\circle*{0.75}}
 \put( 39.20, 55.96){\circle*{0.75}}
 \put( 39.30, 55.87){\circle*{0.75}}
 \put( 39.40, 55.77){\circle*{0.75}}
 \put( 39.50, 55.68){\circle*{0.75}}
 \put( 39.60, 55.59){\circle*{0.75}}
 \put( 39.70, 55.49){\circle*{0.75}}
 \put( 39.80, 55.40){\circle*{0.75}}
 \put( 39.90, 55.31){\circle*{0.75}}
 \put( 40.00, 55.22){\circle*{0.75}}
 \put( 40.10, 55.13){\circle*{0.75}}
 \put( 40.20, 55.04){\circle*{0.75}}
 \put( 40.30, 54.95){\circle*{0.75}}
 \put( 40.40, 54.86){\circle*{0.75}}
 \put( 40.50, 54.78){\circle*{0.75}}
 \put( 40.60, 54.69){\circle*{0.75}}
 \put( 40.70, 54.60){\circle*{0.75}}
 \put( 40.80, 54.51){\circle*{0.75}}
 \put( 40.90, 54.43){\circle*{0.75}}
 \put( 41.00, 54.34){\circle*{0.75}}
 \put( 41.10, 54.26){\circle*{0.75}}
 \put( 41.20, 54.17){\circle*{0.75}}
 \put( 41.30, 54.09){\circle*{0.75}}
 \put( 41.40, 54.01){\circle*{0.75}}
 \put( 41.50, 53.92){\circle*{0.75}}
 \put( 41.60, 53.84){\circle*{0.75}}
 \put( 41.70, 53.76){\circle*{0.75}}
 \put( 41.80, 53.68){\circle*{0.75}}
 \put( 41.90, 53.59){\circle*{0.75}}
 \put( 42.00, 53.51){\circle*{0.75}}
 \put( 42.10, 53.43){\circle*{0.75}}
 \put( 42.20, 53.35){\circle*{0.75}}
 \put( 42.30, 53.27){\circle*{0.75}}
 \put( 42.40, 53.19){\circle*{0.75}}
 \put( 42.50, 53.12){\circle*{0.75}}
 \put( 42.60, 53.04){\circle*{0.75}}
 \put( 42.70, 52.96){\circle*{0.75}}
 \put( 42.80, 52.88){\circle*{0.75}}
 \put( 42.90, 52.81){\circle*{0.75}}
 \put( 43.00, 52.73){\circle*{0.75}}
 \put( 43.10, 52.65){\circle*{0.75}}
 \put( 43.20, 52.58){\circle*{0.75}}
 \put( 43.30, 52.50){\circle*{0.75}}
 \put( 43.40, 52.43){\circle*{0.75}}
 \put( 43.50, 52.35){\circle*{0.75}}
 \put( 43.60, 52.28){\circle*{0.75}}
 \put( 43.70, 52.21){\circle*{0.75}}
 \put( 43.80, 52.13){\circle*{0.75}}
 \put( 43.90, 52.06){\circle*{0.75}}
 \put( 44.00, 51.99){\circle*{0.75}}
 \put( 44.10, 51.91){\circle*{0.75}}
 \put( 44.20, 51.84){\circle*{0.75}}
 \put( 44.30, 51.77){\circle*{0.75}}
 \put( 44.40, 51.70){\circle*{0.75}}
 \put( 44.50, 51.63){\circle*{0.75}}
 \put( 44.60, 51.56){\circle*{0.75}}
 \put( 44.70, 51.49){\circle*{0.75}}
 \put( 44.80, 51.42){\circle*{0.75}}
 \put( 44.90, 51.35){\circle*{0.75}}
 \put( 45.00, 51.28){\circle*{0.75}}
 \put( 45.10, 51.21){\circle*{0.75}}
 \put( 45.20, 51.14){\circle*{0.75}}
 \put( 45.30, 51.08){\circle*{0.75}}
 \put( 45.40, 51.01){\circle*{0.75}}
 \put( 45.50, 50.94){\circle*{0.75}}
 \put( 45.60, 50.87){\circle*{0.75}}
 \put( 45.70, 50.81){\circle*{0.75}}
 \put( 45.80, 50.74){\circle*{0.75}}
 \put( 45.90, 50.67){\circle*{0.75}}
 \put( 46.00, 50.61){\circle*{0.75}}
 \put( 46.10, 50.54){\circle*{0.75}}
 \put( 46.20, 50.48){\circle*{0.75}}
 \put( 46.30, 50.41){\circle*{0.75}}
 \put( 46.40, 50.35){\circle*{0.75}}
 \put( 46.50, 50.29){\circle*{0.75}}
 \put( 46.60, 50.22){\circle*{0.75}}
 \put( 46.70, 50.16){\circle*{0.75}}
 \put( 46.80, 50.10){\circle*{0.75}}
 \put( 46.90, 50.03){\circle*{0.75}}
 \put( 47.00, 49.97){\circle*{0.75}}
 \put( 47.10, 49.91){\circle*{0.75}}
 \put( 47.20, 49.85){\circle*{0.75}}
 \put( 47.30, 49.78){\circle*{0.75}}
 \put( 47.40, 49.72){\circle*{0.75}}
 \put( 47.50, 49.66){\circle*{0.75}}
 \put( 47.60, 49.60){\circle*{0.75}}
 \put( 47.70, 49.54){\circle*{0.75}}
 \put( 47.80, 49.48){\circle*{0.75}}
 \put( 47.90, 49.42){\circle*{0.75}}
 \put( 48.00, 49.36){\circle*{0.75}}
 \put( 48.10, 49.30){\circle*{0.75}}
 \put( 48.20, 49.24){\circle*{0.75}}
 \put( 48.30, 49.18){\circle*{0.75}}
 \put( 48.40, 49.12){\circle*{0.75}}
 \put( 48.50, 49.06){\circle*{0.75}}
 \put( 48.60, 49.01){\circle*{0.75}}
 \put( 48.70, 48.95){\circle*{0.75}}
 \put( 48.80, 48.89){\circle*{0.75}}
 \put( 48.90, 48.83){\circle*{0.75}}
 \put( 49.00, 48.78){\circle*{0.75}}
 \put( 49.10, 48.72){\circle*{0.75}}
 \put( 49.20, 48.66){\circle*{0.75}}
 \put( 49.30, 48.61){\circle*{0.75}}
 \put( 49.40, 48.55){\circle*{0.75}}
 \put( 49.50, 48.49){\circle*{0.75}}
 \put( 49.60, 48.44){\circle*{0.75}}
 \put( 49.70, 48.38){\circle*{0.75}}
 \put( 49.80, 48.33){\circle*{0.75}}
 \put( 49.90, 48.27){\circle*{0.75}}
 \put( 50.00, 48.22){\circle*{0.75}}
 \put( 50.10, 48.16){\circle*{0.75}}
 \put( 50.20, 48.11){\circle*{0.75}}
 \put( 50.30, 48.06){\circle*{0.75}}
 \put( 50.40, 48.00){\circle*{0.75}}
 \put( 50.50, 47.95){\circle*{0.75}}
 \put( 50.60, 47.89){\circle*{0.75}}
 \put( 50.70, 47.84){\circle*{0.75}}
 \put( 50.80, 47.79){\circle*{0.75}}
 \put( 50.90, 47.74){\circle*{0.75}}
 \put( 51.00, 47.68){\circle*{0.75}}
 \put( 51.10, 47.63){\circle*{0.75}}
 \put( 51.20, 47.58){\circle*{0.75}}
 \put( 51.30, 47.53){\circle*{0.75}}
 \put( 51.40, 47.48){\circle*{0.75}}
 \put( 51.50, 47.42){\circle*{0.75}}
 \put( 51.60, 47.37){\circle*{0.75}}
 \put( 51.70, 47.32){\circle*{0.75}}
 \put( 51.80, 47.27){\circle*{0.75}}
 \put( 51.90, 47.22){\circle*{0.75}}
 \put( 52.00, 47.17){\circle*{0.75}}
 \put( 52.10, 47.12){\circle*{0.75}}
 \put( 52.20, 47.07){\circle*{0.75}}
 \put( 52.30, 47.02){\circle*{0.75}}
 \put( 52.40, 46.97){\circle*{0.75}}
 \put( 52.50, 46.92){\circle*{0.75}}
 \put( 52.60, 46.87){\circle*{0.75}}
 \put( 52.70, 46.82){\circle*{0.75}}
 \put( 52.80, 46.78){\circle*{0.75}}
 \put( 52.90, 46.73){\circle*{0.75}}
 \put( 53.00, 46.68){\circle*{0.75}}
 \put( 53.10, 46.63){\circle*{0.75}}
 \put( 53.20, 46.58){\circle*{0.75}}
 \put( 53.30, 46.53){\circle*{0.75}}
 \put( 53.40, 46.49){\circle*{0.75}}
 \put( 53.50, 46.44){\circle*{0.75}}
 \put( 53.60, 46.39){\circle*{0.75}}
 \put( 53.70, 46.35){\circle*{0.75}}
 \put( 53.80, 46.30){\circle*{0.75}}
 \put( 53.90, 46.25){\circle*{0.75}}
 \put( 54.00, 46.21){\circle*{0.75}}
 \put( 54.10, 46.16){\circle*{0.75}}
 \put( 54.20, 46.11){\circle*{0.75}}
 \put( 54.30, 46.07){\circle*{0.75}}
 \put( 54.40, 46.02){\circle*{0.75}}
 \put( 54.50, 45.98){\circle*{0.75}}
 \put( 54.60, 45.93){\circle*{0.75}}
 \put( 54.70, 45.88){\circle*{0.75}}
 \put( 54.80, 45.84){\circle*{0.75}}
 \put( 54.90, 45.79){\circle*{0.75}}
 \put( 55.00, 45.75){\circle*{0.75}}
 \put( 55.10, 45.71){\circle*{0.75}}
 \put( 55.20, 45.66){\circle*{0.75}}
 \put( 55.30, 45.62){\circle*{0.75}}
 \put( 55.40, 45.57){\circle*{0.75}}
 \put( 55.50, 45.53){\circle*{0.75}}
 \put( 55.60, 45.48){\circle*{0.75}}
 \put( 55.70, 45.44){\circle*{0.75}}
 \put( 55.80, 45.40){\circle*{0.75}}
 \put( 55.90, 45.35){\circle*{0.75}}
 \put( 56.00, 45.31){\circle*{0.75}}
 \put( 56.10, 45.27){\circle*{0.75}}
 \put( 56.20, 45.23){\circle*{0.75}}
 \put( 56.30, 45.18){\circle*{0.75}}
 \put( 56.40, 45.14){\circle*{0.75}}
 \put( 56.50, 45.10){\circle*{0.75}}
 \put( 56.60, 45.06){\circle*{0.75}}
 \put( 56.70, 45.01){\circle*{0.75}}
 \put( 56.80, 44.97){\circle*{0.75}}
 \put( 56.90, 44.93){\circle*{0.75}}
 \put( 57.00, 44.89){\circle*{0.75}}
 \put( 57.10, 44.85){\circle*{0.75}}
 \put( 57.20, 44.81){\circle*{0.75}}
 \put( 57.30, 44.76){\circle*{0.75}}
 \put( 57.40, 44.72){\circle*{0.75}}
 \put( 57.50, 44.68){\circle*{0.75}}
 \put( 57.60, 44.64){\circle*{0.75}}
 \put( 57.70, 44.60){\circle*{0.75}}
 \put( 57.80, 44.56){\circle*{0.75}}
 \put( 57.90, 44.52){\circle*{0.75}}
 \put( 58.00, 44.48){\circle*{0.75}}
 \put( 58.10, 44.44){\circle*{0.75}}
 \put( 58.20, 44.40){\circle*{0.75}}
 \put( 58.30, 44.36){\circle*{0.75}}
 \put( 58.40, 44.32){\circle*{0.75}}
 \put( 58.50, 44.28){\circle*{0.75}}
 \put( 58.60, 44.24){\circle*{0.75}}
 \put( 58.70, 44.20){\circle*{0.75}}
 \put( 58.80, 44.16){\circle*{0.75}}
 \put( 58.90, 44.13){\circle*{0.75}}
 \put( 59.00, 44.09){\circle*{0.75}}
 \put( 59.10, 44.05){\circle*{0.75}}
 \put( 59.20, 44.01){\circle*{0.75}}
 \put( 59.30, 43.97){\circle*{0.75}}
 \put( 59.40, 43.93){\circle*{0.75}}
 \put( 59.50, 43.89){\circle*{0.75}}
 \put( 59.60, 43.86){\circle*{0.75}}
 \put( 59.70, 43.82){\circle*{0.75}}
 \put( 59.80, 43.78){\circle*{0.75}}
 \put( 59.90, 43.74){\circle*{0.75}}
 \put( 60.00, 43.71){\circle*{0.75}}
 \put( 60.10, 43.67){\circle*{0.75}}
 \put( 60.20, 43.63){\circle*{0.75}}
 \put( 60.30, 43.59){\circle*{0.75}}
 \put( 60.40, 43.56){\circle*{0.75}}
 \put( 60.50, 43.52){\circle*{0.75}}
 \put( 60.60, 43.48){\circle*{0.75}}
 \put( 60.70, 43.45){\circle*{0.75}}
 \put( 60.80, 43.41){\circle*{0.75}}
 \put( 60.90, 43.37){\circle*{0.75}}
 \put( 61.00, 43.34){\circle*{0.75}}
 \put( 61.10, 43.30){\circle*{0.75}}
 \put( 61.20, 43.26){\circle*{0.75}}
 \put( 61.30, 43.23){\circle*{0.75}}
 \put( 61.40, 43.19){\circle*{0.75}}
 \put( 61.50, 43.16){\circle*{0.75}}
 \put( 61.60, 43.12){\circle*{0.75}}
 \put( 61.70, 43.09){\circle*{0.75}}
 \put( 61.80, 43.05){\circle*{0.75}}
 \put( 61.90, 43.02){\circle*{0.75}}
 \put( 62.00, 42.98){\circle*{0.75}}
 \put( 62.10, 42.95){\circle*{0.75}}
 \put( 62.20, 42.91){\circle*{0.75}}
 \put( 62.30, 42.88){\circle*{0.75}}
 \put( 62.40, 42.84){\circle*{0.75}}
 \put( 62.50, 42.81){\circle*{0.75}}
 \put( 62.60, 42.77){\circle*{0.75}}
 \put( 62.70, 42.74){\circle*{0.75}}
 \put( 62.80, 42.70){\circle*{0.75}}
 \put( 62.90, 42.67){\circle*{0.75}}
 \put( 63.00, 42.64){\circle*{0.75}}
 \put( 63.10, 42.60){\circle*{0.75}}
 \put( 63.20, 42.57){\circle*{0.75}}
 \put( 63.30, 42.53){\circle*{0.75}}
 \put( 63.40, 42.50){\circle*{0.75}}
 \put( 63.50, 42.47){\circle*{0.75}}
 \put( 63.60, 42.43){\circle*{0.75}}
 \put( 63.70, 42.40){\circle*{0.75}}
 \put( 63.80, 42.37){\circle*{0.75}}
 \put( 63.90, 42.33){\circle*{0.75}}
 \put( 64.00, 42.30){\circle*{0.75}}
 \put( 64.10, 42.27){\circle*{0.75}}
 \put( 64.20, 42.23){\circle*{0.75}}
 \put( 64.30, 42.20){\circle*{0.75}}
 \put( 64.40, 42.17){\circle*{0.75}}
 \put( 64.50, 42.14){\circle*{0.75}}
 \put( 64.60, 42.10){\circle*{0.75}}
 \put( 64.70, 42.07){\circle*{0.75}}
 \put( 64.80, 42.04){\circle*{0.75}}
 \put( 64.90, 42.01){\circle*{0.75}}
 \put( 65.00, 41.98){\circle*{0.75}}
 \put( 65.10, 41.94){\circle*{0.75}}
 \put( 65.20, 41.91){\circle*{0.75}}
 \put( 65.30, 41.88){\circle*{0.75}}
 \put( 65.40, 41.85){\circle*{0.75}}
 \put( 65.50, 41.82){\circle*{0.75}}
 \put( 65.60, 41.79){\circle*{0.75}}
 \put( 65.70, 41.75){\circle*{0.75}}
 \put( 65.80, 41.72){\circle*{0.75}}
 \put( 65.90, 41.69){\circle*{0.75}}
 \put( 66.00, 41.66){\circle*{0.75}}
 \put( 66.10, 41.63){\circle*{0.75}}
 \put( 66.20, 41.60){\circle*{0.75}}
 \put( 66.30, 41.57){\circle*{0.75}}
 \put( 66.40, 41.54){\circle*{0.75}}
 \put( 66.50, 41.51){\circle*{0.75}}
 \put( 66.60, 41.48){\circle*{0.75}}
 \put( 66.70, 41.45){\circle*{0.75}}
 \put( 66.80, 41.42){\circle*{0.75}}
 \put( 66.90, 41.39){\circle*{0.75}}
 \put( 67.00, 41.35){\circle*{0.75}}
 \put( 67.10, 41.32){\circle*{0.75}}
 \put( 67.20, 41.29){\circle*{0.75}}
 \put( 67.30, 41.26){\circle*{0.75}}
 \put( 67.40, 41.23){\circle*{0.75}}
 \put( 67.50, 41.21){\circle*{0.75}}
 \put( 67.60, 41.18){\circle*{0.75}}
 \put( 67.70, 41.15){\circle*{0.75}}
 \put( 67.80, 41.12){\circle*{0.75}}
 \put( 67.90, 41.09){\circle*{0.75}}
 \put( 68.00, 41.06){\circle*{0.75}}
 \put( 68.10, 41.03){\circle*{0.75}}
 \put( 68.20, 41.00){\circle*{0.75}}
 \put( 68.30, 40.97){\circle*{0.75}}
 \put( 68.40, 40.94){\circle*{0.75}}
 \put( 68.50, 40.91){\circle*{0.75}}
 \put( 68.60, 40.88){\circle*{0.75}}
 \put( 68.70, 40.85){\circle*{0.75}}
 \put( 68.80, 40.83){\circle*{0.75}}
 \put( 68.90, 40.80){\circle*{0.75}}
 \put( 69.00, 40.77){\circle*{0.75}}
 \put( 69.10, 40.74){\circle*{0.75}}
 \put( 69.20, 40.71){\circle*{0.75}}
 \put( 69.30, 40.68){\circle*{0.75}}
 \put( 69.40, 40.66){\circle*{0.75}}
 \put( 69.50, 40.63){\circle*{0.75}}
 \put( 69.60, 40.60){\circle*{0.75}}
 \put( 69.70, 40.57){\circle*{0.75}}
 \put( 69.80, 40.54){\circle*{0.75}}
 \put( 69.90, 40.52){\circle*{0.75}}
 \put( 70.00, 40.49){\circle*{0.75}}
 \put( 70.10, 40.46){\circle*{0.75}}
 \put( 70.20, 40.43){\circle*{0.75}}
 \put( 70.30, 40.40){\circle*{0.75}}
 \put( 70.40, 40.38){\circle*{0.75}}
 \put( 70.50, 40.35){\circle*{0.75}}
 \put( 70.60, 40.32){\circle*{0.75}}
 \put( 70.70, 40.30){\circle*{0.75}}
 \put( 70.80, 40.27){\circle*{0.75}}
 \put( 70.90, 40.24){\circle*{0.75}}
 \put( 71.00, 40.21){\circle*{0.75}}
 \put( 71.10, 40.19){\circle*{0.75}}
 \put( 71.20, 40.16){\circle*{0.75}}
 \put( 71.30, 40.13){\circle*{0.75}}
 \put( 71.40, 40.11){\circle*{0.75}}
 \put( 71.50, 40.08){\circle*{0.75}}
 \put( 71.60, 40.05){\circle*{0.75}}
 \put( 71.70, 40.03){\circle*{0.75}}
 \put( 71.80, 40.00){\circle*{0.75}}
 \put( 71.90, 39.97){\circle*{0.75}}
 \put( 72.00, 39.95){\circle*{0.75}}
 \put( 72.10, 39.92){\circle*{0.75}}
 \put( 72.20, 39.90){\circle*{0.75}}
 \put( 72.30, 39.87){\circle*{0.75}}
 \put( 72.40, 39.84){\circle*{0.75}}
 \put( 72.50, 39.82){\circle*{0.75}}
 \put( 72.60, 39.79){\circle*{0.75}}
 \put( 72.70, 39.77){\circle*{0.75}}
 \put( 72.80, 39.74){\circle*{0.75}}
 \put( 72.90, 39.71){\circle*{0.75}}
 \put( 73.00, 39.69){\circle*{0.75}}
 \put( 73.10, 39.66){\circle*{0.75}}
 \put( 73.20, 39.64){\circle*{0.75}}
 \put( 73.30, 39.61){\circle*{0.75}}
 \put( 73.40, 39.59){\circle*{0.75}}
 \put( 73.50, 39.56){\circle*{0.75}}
 \put( 73.60, 39.54){\circle*{0.75}}
 \put( 73.70, 39.51){\circle*{0.75}}
 \put( 73.80, 39.49){\circle*{0.75}}
 \put( 73.90, 39.46){\circle*{0.75}}
 \put( 74.00, 39.44){\circle*{0.75}}
 \put( 74.10, 39.41){\circle*{0.75}}
 \put( 74.20, 39.39){\circle*{0.75}}
 \put( 74.30, 39.36){\circle*{0.75}}
 \put( 74.40, 39.34){\circle*{0.75}}
 \put( 74.50, 39.31){\circle*{0.75}}
 \put( 74.60, 39.29){\circle*{0.75}}
 \put( 74.70, 39.26){\circle*{0.75}}
 \put( 74.80, 39.24){\circle*{0.75}}
 \put( 74.90, 39.21){\circle*{0.75}}
 \put( 75.00, 39.19){\circle*{0.75}}
 \put( 75.10, 39.17){\circle*{0.75}}
 \put( 75.20, 39.14){\circle*{0.75}}
 \put( 75.30, 39.12){\circle*{0.75}}
 \put( 75.40, 39.09){\circle*{0.75}}
 \put( 75.50, 39.07){\circle*{0.75}}
 \put( 75.60, 39.04){\circle*{0.75}}
 \put( 75.70, 39.02){\circle*{0.75}}
 \put( 75.80, 39.00){\circle*{0.75}}
 \put( 75.90, 38.97){\circle*{0.75}}
 \put( 76.00, 38.95){\circle*{0.75}}
 \put( 76.10, 38.93){\circle*{0.75}}
 \put( 76.20, 38.90){\circle*{0.75}}
 \put( 76.30, 38.88){\circle*{0.75}}
 \put( 76.40, 38.86){\circle*{0.75}}
 \put( 76.50, 38.83){\circle*{0.75}}
 \put( 76.60, 38.81){\circle*{0.75}}
 \put( 76.70, 38.78){\circle*{0.75}}
 \put( 76.80, 38.76){\circle*{0.75}}
 \put( 76.90, 38.74){\circle*{0.75}}
 \put( 77.00, 38.72){\circle*{0.75}}
 \put( 77.10, 38.69){\circle*{0.75}}
 \put( 77.20, 38.67){\circle*{0.75}}
 \put( 77.30, 38.65){\circle*{0.75}}
 \put( 77.40, 38.62){\circle*{0.75}}
 \put( 77.50, 38.60){\circle*{0.75}}
 \put( 77.60, 38.58){\circle*{0.75}}
 \put( 77.70, 38.55){\circle*{0.75}}
 \put( 77.80, 38.53){\circle*{0.75}}
 \put( 77.90, 38.51){\circle*{0.75}}
 \put( 78.00, 38.49){\circle*{0.75}}
 \put( 78.10, 38.46){\circle*{0.75}}
 \put( 78.20, 38.44){\circle*{0.75}}
 \put( 78.30, 38.42){\circle*{0.75}}
 \put( 78.40, 38.40){\circle*{0.75}}
 \put( 78.50, 38.37){\circle*{0.75}}
 \put( 78.60, 38.35){\circle*{0.75}}
 \put( 78.70, 38.33){\circle*{0.75}}
 \put( 78.80, 38.31){\circle*{0.75}}
 \put( 78.90, 38.28){\circle*{0.75}}
 \put( 79.00, 38.26){\circle*{0.75}}
 \put( 79.10, 38.24){\circle*{0.75}}
 \put( 79.20, 38.22){\circle*{0.75}}
 \put( 79.30, 38.20){\circle*{0.75}}
 \put( 79.40, 38.17){\circle*{0.75}}
 \put( 79.50, 38.15){\circle*{0.75}}
 \put( 79.60, 38.13){\circle*{0.75}}
 \put( 79.70, 38.11){\circle*{0.75}}
 \put( 79.80, 38.09){\circle*{0.75}}
 \put( 79.90, 38.07){\circle*{0.75}}
 \put( 80.00, 38.04){\circle*{0.75}}

\end{picture}
\end{center}
\caption{The calculated dependence of $nS$ - bottomonium leptonic
constants and the experimental values of $f_{\Upsilon(nS)}$.}
\label{fh1}
\end{figure}
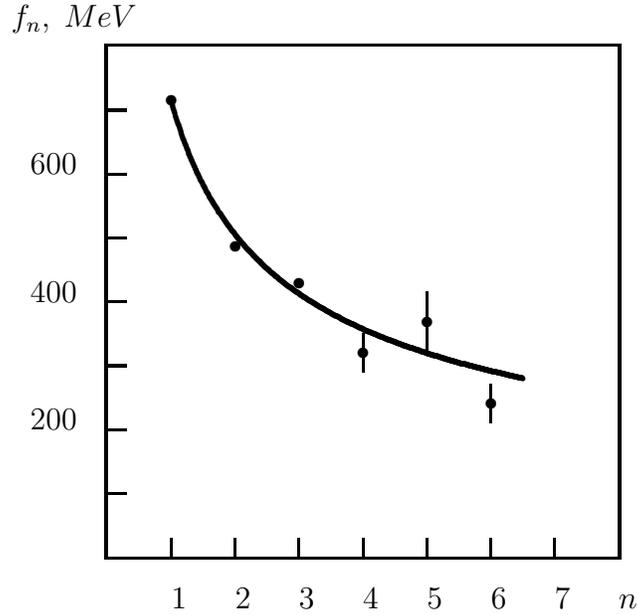
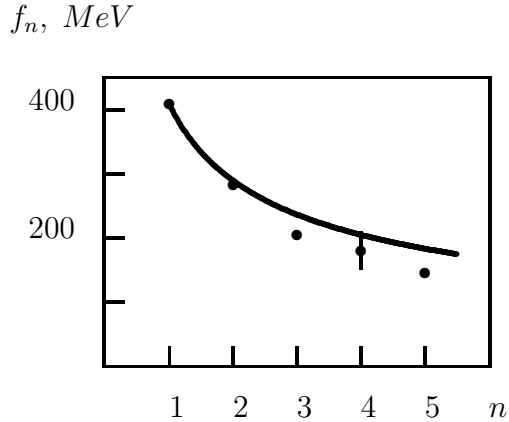
\begin{figure}[t]
\begin{center}
\begin{picture}(100,60)
\put(15,10){\framebox(60,45)}
\put(15,30){\line(1,0){3}}
\put(3,30){$200$}
\put(15,50){\line(1,0){3}}
\put(3,50){$400$}
\put(15,20){\line(1,0){3}}
\put(15,40){\line(1,0){3}}

\put(0,63){$f_n,\;MeV$}

\put(25,10){\line(0,1){3}}
\put(35,10){\line(0,1){3}}
\put(45,10){\line(0,1){3}}
\put(55,10){\line(0,1){3}}
\put(65,10){\line(0,1){3}}
\put(25,2){$1$}
\put(35,2){$2$}
\put(45,2){$3$}
\put(55,2){$4$}
\put(65,2){$5$}
\put(75,2){$n$}

\put(25,51.0){\circle*{1.6}}
\put(35,38.3){\circle*{1.6}}
\put(45,30.5){\circle*{1.6}}
\put(55,28.0){\circle*{1.6}}
\put(65,24.5){\circle*{1.6}}
\put(55,25.0){\line(0,1){6}}

 \put( 25.10, 50.80){\circle*{0.75}}
 \put( 25.20, 50.60){\circle*{0.75}}
 \put( 25.30, 50.40){\circle*{0.75}}
 \put( 25.40, 50.20){\circle*{0.75}}
 \put( 25.50, 50.01){\circle*{0.75}}
 \put( 25.60, 49.82){\circle*{0.75}}
 \put( 25.70, 49.64){\circle*{0.75}}
 \put( 25.80, 49.45){\circle*{0.75}}
 \put( 25.90, 49.27){\circle*{0.75}}
 \put( 26.00, 49.09){\circle*{0.75}}
 \put( 26.10, 48.92){\circle*{0.75}}
 \put( 26.20, 48.74){\circle*{0.75}}
 \put( 26.30, 48.57){\circle*{0.75}}
 \put( 26.40, 48.40){\circle*{0.75}}
 \put( 26.50, 48.23){\circle*{0.75}}
 \put( 26.60, 48.07){\circle*{0.75}}
 \put( 26.70, 47.90){\circle*{0.75}}
 \put( 26.80, 47.74){\circle*{0.75}}
 \put( 26.90, 47.58){\circle*{0.75}}
 \put( 27.00, 47.43){\circle*{0.75}}
 \put( 27.10, 47.27){\circle*{0.75}}
 \put( 27.20, 47.12){\circle*{0.75}}
 \put( 27.30, 46.97){\circle*{0.75}}
 \put( 27.40, 46.82){\circle*{0.75}}
 \put( 27.50, 46.67){\circle*{0.75}}
 \put( 27.60, 46.53){\circle*{0.75}}
 \put( 27.70, 46.38){\circle*{0.75}}
 \put( 27.80, 46.24){\circle*{0.75}}
 \put( 27.90, 46.10){\circle*{0.75}}
 \put( 28.00, 45.96){\circle*{0.75}}
 \put( 28.10, 45.82){\circle*{0.75}}
 \put( 28.20, 45.69){\circle*{0.75}}
 \put( 28.30, 45.55){\circle*{0.75}}
 \put( 28.40, 45.42){\circle*{0.75}}
 \put( 28.50, 45.29){\circle*{0.75}}
 \put( 28.60, 45.16){\circle*{0.75}}
 \put( 28.70, 45.03){\circle*{0.75}}
 \put( 28.80, 44.90){\circle*{0.75}}
 \put( 28.90, 44.78){\circle*{0.75}}
 \put( 29.00, 44.65){\circle*{0.75}}
 \put( 29.10, 44.53){\circle*{0.75}}
 \put( 29.20, 44.41){\circle*{0.75}}
 \put( 29.30, 44.29){\circle*{0.75}}
 \put( 29.40, 44.17){\circle*{0.75}}
 \put( 29.50, 44.05){\circle*{0.75}}
 \put( 29.60, 43.93){\circle*{0.75}}
 \put( 29.70, 43.82){\circle*{0.75}}
 \put( 29.80, 43.70){\circle*{0.75}}
 \put( 29.90, 43.59){\circle*{0.75}}
 \put( 30.00, 43.48){\circle*{0.75}}
 \put( 30.10, 43.37){\circle*{0.75}}
 \put( 30.20, 43.26){\circle*{0.75}}
 \put( 30.30, 43.15){\circle*{0.75}}
 \put( 30.40, 43.04){\circle*{0.75}}
 \put( 30.50, 42.93){\circle*{0.75}}
 \put( 30.60, 42.83){\circle*{0.75}}
 \put( 30.70, 42.72){\circle*{0.75}}
 \put( 30.80, 42.62){\circle*{0.75}}
 \put( 30.90, 42.52){\circle*{0.75}}
 \put( 31.00, 42.41){\circle*{0.75}}
 \put( 31.10, 42.31){\circle*{0.75}}
 \put( 31.20, 42.21){\circle*{0.75}}
 \put( 31.30, 42.11){\circle*{0.75}}
 \put( 31.40, 42.02){\circle*{0.75}}
 \put( 31.50, 41.92){\circle*{0.75}}
 \put( 31.60, 41.82){\circle*{0.75}}
 \put( 31.70, 41.73){\circle*{0.75}}
 \put( 31.80, 41.63){\circle*{0.75}}
 \put( 31.90, 41.54){\circle*{0.75}}
 \put( 32.00, 41.45){\circle*{0.75}}
 \put( 32.10, 41.35){\circle*{0.75}}
 \put( 32.20, 41.26){\circle*{0.75}}
 \put( 32.30, 41.17){\circle*{0.75}}
 \put( 32.40, 41.08){\circle*{0.75}}
 \put( 32.50, 40.99){\circle*{0.75}}
 \put( 32.60, 40.90){\circle*{0.75}}
 \put( 32.70, 40.82){\circle*{0.75}}
 \put( 32.80, 40.73){\circle*{0.75}}
 \put( 32.90, 40.64){\circle*{0.75}}
 \put( 33.00, 40.56){\circle*{0.75}}
 \put( 33.10, 40.48){\circle*{0.75}}
 \put( 33.20, 40.39){\circle*{0.75}}
 \put( 33.30, 40.31){\circle*{0.75}}
 \put( 33.40, 40.23){\circle*{0.75}}
 \put( 33.50, 40.14){\circle*{0.75}}
 \put( 33.60, 40.06){\circle*{0.75}}
 \put( 33.70, 39.98){\circle*{0.75}}
 \put( 33.80, 39.90){\circle*{0.75}}
 \put( 33.90, 39.82){\circle*{0.75}}
 \put( 34.00, 39.74){\circle*{0.75}}
 \put( 34.10, 39.67){\circle*{0.75}}
 \put( 34.20, 39.59){\circle*{0.75}}
 \put( 34.30, 39.51){\circle*{0.75}}
 \put( 34.40, 39.44){\circle*{0.75}}
 \put( 34.50, 39.36){\circle*{0.75}}
 \put( 34.60, 39.29){\circle*{0.75}}
 \put( 34.70, 39.21){\circle*{0.75}}
 \put( 34.80, 39.14){\circle*{0.75}}
 \put( 34.90, 39.06){\circle*{0.75}}
 \put( 35.00, 38.99){\circle*{0.75}}
 \put( 35.10, 38.92){\circle*{0.75}}
 \put( 35.20, 38.85){\circle*{0.75}}
 \put( 35.30, 38.78){\circle*{0.75}}
 \put( 35.40, 38.71){\circle*{0.75}}
 \put( 35.50, 38.64){\circle*{0.75}}
 \put( 35.60, 38.57){\circle*{0.75}}
 \put( 35.70, 38.50){\circle*{0.75}}
 \put( 35.80, 38.43){\circle*{0.75}}
 \put( 35.90, 38.36){\circle*{0.75}}
 \put( 36.00, 38.29){\circle*{0.75}}
 \put( 36.10, 38.23){\circle*{0.75}}
 \put( 36.20, 38.16){\circle*{0.75}}
 \put( 36.30, 38.09){\circle*{0.75}}
 \put( 36.40, 38.03){\circle*{0.75}}
 \put( 36.50, 37.96){\circle*{0.75}}
 \put( 36.60, 37.90){\circle*{0.75}}
 \put( 36.70, 37.83){\circle*{0.75}}
 \put( 36.80, 37.77){\circle*{0.75}}
 \put( 36.90, 37.71){\circle*{0.75}}
 \put( 37.00, 37.64){\circle*{0.75}}
 \put( 37.10, 37.58){\circle*{0.75}}
 \put( 37.20, 37.52){\circle*{0.75}}
 \put( 37.30, 37.46){\circle*{0.75}}
 \put( 37.40, 37.39){\circle*{0.75}}
 \put( 37.50, 37.33){\circle*{0.75}}
 \put( 37.60, 37.27){\circle*{0.75}}
 \put( 37.70, 37.21){\circle*{0.75}}
 \put( 37.80, 37.15){\circle*{0.75}}
 \put( 37.90, 37.09){\circle*{0.75}}
 \put( 38.00, 37.03){\circle*{0.75}}
 \put( 38.10, 36.98){\circle*{0.75}}
 \put( 38.20, 36.92){\circle*{0.75}}
 \put( 38.30, 36.86){\circle*{0.75}}
 \put( 38.40, 36.80){\circle*{0.75}}
 \put( 38.50, 36.75){\circle*{0.75}}
 \put( 38.60, 36.69){\circle*{0.75}}
 \put( 38.70, 36.63){\circle*{0.75}}
 \put( 38.80, 36.58){\circle*{0.75}}
 \put( 38.90, 36.52){\circle*{0.75}}
 \put( 39.00, 36.47){\circle*{0.75}}
 \put( 39.10, 36.41){\circle*{0.75}}
 \put( 39.20, 36.36){\circle*{0.75}}
 \put( 39.30, 36.30){\circle*{0.75}}
 \put( 39.40, 36.25){\circle*{0.75}}
 \put( 39.50, 36.19){\circle*{0.75}}
 \put( 39.60, 36.14){\circle*{0.75}}
 \put( 39.70, 36.09){\circle*{0.75}}
 \put( 39.80, 36.04){\circle*{0.75}}
 \put( 39.90, 35.98){\circle*{0.75}}
 \put( 40.00, 35.93){\circle*{0.75}}
 \put( 40.10, 35.88){\circle*{0.75}}
 \put( 40.20, 35.83){\circle*{0.75}}
 \put( 40.30, 35.78){\circle*{0.75}}
 \put( 40.40, 35.73){\circle*{0.75}}
 \put( 40.50, 35.68){\circle*{0.75}}
 \put( 40.60, 35.63){\circle*{0.75}}
 \put( 40.70, 35.58){\circle*{0.75}}
 \put( 40.80, 35.53){\circle*{0.75}}
 \put( 40.90, 35.48){\circle*{0.75}}
 \put( 41.00, 35.43){\circle*{0.75}}
 \put( 41.10, 35.38){\circle*{0.75}}
 \put( 41.20, 35.33){\circle*{0.75}}
 \put( 41.30, 35.28){\circle*{0.75}}
 \put( 41.40, 35.23){\circle*{0.75}}
 \put( 41.50, 35.19){\circle*{0.75}}
 \put( 41.60, 35.14){\circle*{0.75}}
 \put( 41.70, 35.09){\circle*{0.75}}
 \put( 41.80, 35.04){\circle*{0.75}}
 \put( 41.90, 35.00){\circle*{0.75}}
 \put( 42.00, 34.95){\circle*{0.75}}
 \put( 42.10, 34.91){\circle*{0.75}}
 \put( 42.20, 34.86){\circle*{0.75}}
 \put( 42.30, 34.81){\circle*{0.75}}
 \put( 42.40, 34.77){\circle*{0.75}}
 \put( 42.50, 34.72){\circle*{0.75}}
 \put( 42.60, 34.68){\circle*{0.75}}
 \put( 42.70, 34.63){\circle*{0.75}}
 \put( 42.80, 34.59){\circle*{0.75}}
 \put( 42.90, 34.55){\circle*{0.75}}
 \put( 43.00, 34.50){\circle*{0.75}}
 \put( 43.10, 34.46){\circle*{0.75}}
 \put( 43.20, 34.42){\circle*{0.75}}
 \put( 43.30, 34.37){\circle*{0.75}}
 \put( 43.40, 34.33){\circle*{0.75}}
 \put( 43.50, 34.29){\circle*{0.75}}
 \put( 43.60, 34.24){\circle*{0.75}}
 \put( 43.70, 34.20){\circle*{0.75}}
 \put( 43.80, 34.16){\circle*{0.75}}
 \put( 43.90, 34.12){\circle*{0.75}}
 \put( 44.00, 34.08){\circle*{0.75}}
 \put( 44.10, 34.03){\circle*{0.75}}
 \put( 44.20, 33.99){\circle*{0.75}}
 \put( 44.30, 33.95){\circle*{0.75}}
 \put( 44.40, 33.91){\circle*{0.75}}
 \put( 44.50, 33.87){\circle*{0.75}}
 \put( 44.60, 33.83){\circle*{0.75}}
 \put( 44.70, 33.79){\circle*{0.75}}
 \put( 44.80, 33.75){\circle*{0.75}}
 \put( 44.90, 33.71){\circle*{0.75}}
 \put( 45.00, 33.67){\circle*{0.75}}
 \put( 45.10, 33.63){\circle*{0.75}}
 \put( 45.20, 33.59){\circle*{0.75}}
 \put( 45.30, 33.55){\circle*{0.75}}
 \put( 45.40, 33.52){\circle*{0.75}}
 \put( 45.50, 33.48){\circle*{0.75}}
 \put( 45.60, 33.44){\circle*{0.75}}
 \put( 45.70, 33.40){\circle*{0.75}}
 \put( 45.80, 33.36){\circle*{0.75}}
 \put( 45.90, 33.32){\circle*{0.75}}
 \put( 46.00, 33.29){\circle*{0.75}}
 \put( 46.10, 33.25){\circle*{0.75}}
 \put( 46.20, 33.21){\circle*{0.75}}
 \put( 46.30, 33.17){\circle*{0.75}}
 \put( 46.40, 33.14){\circle*{0.75}}
 \put( 46.50, 33.10){\circle*{0.75}}
 \put( 46.60, 33.06){\circle*{0.75}}
 \put( 46.70, 33.03){\circle*{0.75}}
 \put( 46.80, 32.99){\circle*{0.75}}
 \put( 46.90, 32.96){\circle*{0.75}}
 \put( 47.00, 32.92){\circle*{0.75}}
 \put( 47.10, 32.88){\circle*{0.75}}
 \put( 47.20, 32.85){\circle*{0.75}}
 \put( 47.30, 32.81){\circle*{0.75}}
 \put( 47.40, 32.78){\circle*{0.75}}
 \put( 47.50, 32.74){\circle*{0.75}}
 \put( 47.60, 32.71){\circle*{0.75}}
 \put( 47.70, 32.67){\circle*{0.75}}
 \put( 47.80, 32.64){\circle*{0.75}}
 \put( 47.90, 32.60){\circle*{0.75}}
 \put( 48.00, 32.57){\circle*{0.75}}
 \put( 48.10, 32.54){\circle*{0.75}}
 \put( 48.20, 32.50){\circle*{0.75}}
 \put( 48.30, 32.47){\circle*{0.75}}
 \put( 48.40, 32.43){\circle*{0.75}}
 \put( 48.50, 32.40){\circle*{0.75}}
 \put( 48.60, 32.37){\circle*{0.75}}
 \put( 48.70, 32.33){\circle*{0.75}}
 \put( 48.80, 32.30){\circle*{0.75}}
 \put( 48.90, 32.27){\circle*{0.75}}
 \put( 49.00, 32.24){\circle*{0.75}}
 \put( 49.10, 32.20){\circle*{0.75}}
 \put( 49.20, 32.17){\circle*{0.75}}
 \put( 49.30, 32.14){\circle*{0.75}}
 \put( 49.40, 32.11){\circle*{0.75}}
 \put( 49.50, 32.07){\circle*{0.75}}
 \put( 49.60, 32.04){\circle*{0.75}}
 \put( 49.70, 32.01){\circle*{0.75}}
 \put( 49.80, 31.98){\circle*{0.75}}
 \put( 49.90, 31.95){\circle*{0.75}}
 \put( 50.00, 31.92){\circle*{0.75}}
 \put( 50.10, 31.88){\circle*{0.75}}
 \put( 50.20, 31.85){\circle*{0.75}}
 \put( 50.30, 31.82){\circle*{0.75}}
 \put( 50.40, 31.79){\circle*{0.75}}
 \put( 50.50, 31.76){\circle*{0.75}}
 \put( 50.60, 31.73){\circle*{0.75}}
 \put( 50.70, 31.70){\circle*{0.75}}
 \put( 50.80, 31.67){\circle*{0.75}}
 \put( 50.90, 31.64){\circle*{0.75}}
 \put( 51.00, 31.61){\circle*{0.75}}
 \put( 51.10, 31.58){\circle*{0.75}}
 \put( 51.20, 31.55){\circle*{0.75}}
 \put( 51.30, 31.52){\circle*{0.75}}
 \put( 51.40, 31.49){\circle*{0.75}}
 \put( 51.50, 31.46){\circle*{0.75}}
 \put( 51.60, 31.43){\circle*{0.75}}
 \put( 51.70, 31.40){\circle*{0.75}}
 \put( 51.80, 31.37){\circle*{0.75}}
 \put( 51.90, 31.34){\circle*{0.75}}
 \put( 52.00, 31.31){\circle*{0.75}}
 \put( 52.10, 31.29){\circle*{0.75}}
 \put( 52.20, 31.26){\circle*{0.75}}
 \put( 52.30, 31.23){\circle*{0.75}}
 \put( 52.40, 31.20){\circle*{0.75}}
 \put( 52.50, 31.17){\circle*{0.75}}
 \put( 52.60, 31.14){\circle*{0.75}}
 \put( 52.70, 31.12){\circle*{0.75}}
 \put( 52.80, 31.09){\circle*{0.75}}
 \put( 52.90, 31.06){\circle*{0.75}}
 \put( 53.00, 31.03){\circle*{0.75}}
 \put( 53.10, 31.00){\circle*{0.75}}
 \put( 53.20, 30.98){\circle*{0.75}}
 \put( 53.30, 30.95){\circle*{0.75}}
 \put( 53.40, 30.92){\circle*{0.75}}
 \put( 53.50, 30.90){\circle*{0.75}}
 \put( 53.60, 30.87){\circle*{0.75}}
 \put( 53.70, 30.84){\circle*{0.75}}
 \put( 53.80, 30.81){\circle*{0.75}}
 \put( 53.90, 30.79){\circle*{0.75}}
 \put( 54.00, 30.76){\circle*{0.75}}
 \put( 54.10, 30.73){\circle*{0.75}}
 \put( 54.20, 30.71){\circle*{0.75}}
 \put( 54.30, 30.68){\circle*{0.75}}
 \put( 54.40, 30.66){\circle*{0.75}}
 \put( 54.50, 30.63){\circle*{0.75}}
 \put( 54.60, 30.60){\circle*{0.75}}
 \put( 54.70, 30.58){\circle*{0.75}}
 \put( 54.80, 30.55){\circle*{0.75}}
 \put( 54.90, 30.53){\circle*{0.75}}
 \put( 55.00, 30.50){\circle*{0.75}}
 \put( 55.10, 30.47){\circle*{0.75}}
 \put( 55.20, 30.45){\circle*{0.75}}
 \put( 55.30, 30.42){\circle*{0.75}}
 \put( 55.40, 30.40){\circle*{0.75}}
 \put( 55.50, 30.37){\circle*{0.75}}
 \put( 55.60, 30.35){\circle*{0.75}}
 \put( 55.70, 30.32){\circle*{0.75}}
 \put( 55.80, 30.30){\circle*{0.75}}
 \put( 55.90, 30.27){\circle*{0.75}}
 \put( 56.00, 30.25){\circle*{0.75}}
 \put( 56.10, 30.22){\circle*{0.75}}
 \put( 56.20, 30.20){\circle*{0.75}}
 \put( 56.30, 30.17){\circle*{0.75}}
 \put( 56.40, 30.15){\circle*{0.75}}
 \put( 56.50, 30.13){\circle*{0.75}}
 \put( 56.60, 30.10){\circle*{0.75}}
 \put( 56.70, 30.08){\circle*{0.75}}
 \put( 56.80, 30.05){\circle*{0.75}}
 \put( 56.90, 30.03){\circle*{0.75}}
 \put( 57.00, 30.01){\circle*{0.75}}
 \put( 57.10, 29.98){\circle*{0.75}}
 \put( 57.20, 29.96){\circle*{0.75}}
 \put( 57.30, 29.93){\circle*{0.75}}
 \put( 57.40, 29.91){\circle*{0.75}}
 \put( 57.50, 29.89){\circle*{0.75}}
 \put( 57.60, 29.86){\circle*{0.75}}
 \put( 57.70, 29.84){\circle*{0.75}}
 \put( 57.80, 29.82){\circle*{0.75}}
 \put( 57.90, 29.79){\circle*{0.75}}
 \put( 58.00, 29.77){\circle*{0.75}}
 \put( 58.10, 29.75){\circle*{0.75}}
 \put( 58.20, 29.73){\circle*{0.75}}
 \put( 58.30, 29.70){\circle*{0.75}}
 \put( 58.40, 29.68){\circle*{0.75}}
 \put( 58.50, 29.66){\circle*{0.75}}
 \put( 58.60, 29.64){\circle*{0.75}}
 \put( 58.70, 29.61){\circle*{0.75}}
 \put( 58.80, 29.59){\circle*{0.75}}
 \put( 58.90, 29.57){\circle*{0.75}}
 \put( 59.00, 29.55){\circle*{0.75}}
 \put( 59.10, 29.52){\circle*{0.75}}
 \put( 59.20, 29.50){\circle*{0.75}}
 \put( 59.30, 29.48){\circle*{0.75}}
 \put( 59.40, 29.46){\circle*{0.75}}
 \put( 59.50, 29.44){\circle*{0.75}}
 \put( 59.60, 29.41){\circle*{0.75}}
 \put( 59.70, 29.39){\circle*{0.75}}
 \put( 59.80, 29.37){\circle*{0.75}}
 \put( 59.90, 29.35){\circle*{0.75}}
 \put( 60.00, 29.33){\circle*{0.75}}
 \put( 60.10, 29.31){\circle*{0.75}}
 \put( 60.20, 29.28){\circle*{0.75}}
 \put( 60.30, 29.26){\circle*{0.75}}
 \put( 60.40, 29.24){\circle*{0.75}}
 \put( 60.50, 29.22){\circle*{0.75}}
 \put( 60.60, 29.20){\circle*{0.75}}
 \put( 60.70, 29.18){\circle*{0.75}}
 \put( 60.80, 29.16){\circle*{0.75}}
 \put( 60.90, 29.14){\circle*{0.75}}
 \put( 61.00, 29.12){\circle*{0.75}}
 \put( 61.10, 29.10){\circle*{0.75}}
 \put( 61.20, 29.07){\circle*{0.75}}
 \put( 61.30, 29.05){\circle*{0.75}}
 \put( 61.40, 29.03){\circle*{0.75}}
 \put( 61.50, 29.01){\circle*{0.75}}
 \put( 61.60, 28.99){\circle*{0.75}}
 \put( 61.70, 28.97){\circle*{0.75}}
 \put( 61.80, 28.95){\circle*{0.75}}
 \put( 61.90, 28.93){\circle*{0.75}}
 \put( 62.00, 28.91){\circle*{0.75}}
 \put( 62.10, 28.89){\circle*{0.75}}
 \put( 62.20, 28.87){\circle*{0.75}}
 \put( 62.30, 28.85){\circle*{0.75}}
 \put( 62.40, 28.83){\circle*{0.75}}
 \put( 62.50, 28.81){\circle*{0.75}}
 \put( 62.60, 28.79){\circle*{0.75}}
 \put( 62.70, 28.77){\circle*{0.75}}
 \put( 62.80, 28.75){\circle*{0.75}}
 \put( 62.90, 28.73){\circle*{0.75}}
 \put( 63.00, 28.71){\circle*{0.75}}
 \put( 63.10, 28.69){\circle*{0.75}}
 \put( 63.20, 28.67){\circle*{0.75}}
 \put( 63.30, 28.66){\circle*{0.75}}
 \put( 63.40, 28.64){\circle*{0.75}}
 \put( 63.50, 28.62){\circle*{0.75}}
 \put( 63.60, 28.60){\circle*{0.75}}
 \put( 63.70, 28.58){\circle*{0.75}}
 \put( 63.80, 28.56){\circle*{0.75}}
 \put( 63.90, 28.54){\circle*{0.75}}
 \put( 64.00, 28.52){\circle*{0.75}}
 \put( 64.10, 28.50){\circle*{0.75}}
 \put( 64.20, 28.48){\circle*{0.75}}
 \put( 64.30, 28.47){\circle*{0.75}}
 \put( 64.40, 28.45){\circle*{0.75}}
 \put( 64.50, 28.43){\circle*{0.75}}
 \put( 64.60, 28.41){\circle*{0.75}}
 \put( 64.70, 28.39){\circle*{0.75}}
 \put( 64.80, 28.37){\circle*{0.75}}
 \put( 64.90, 28.35){\circle*{0.75}}
 \put( 65.00, 28.34){\circle*{0.75}}
 \put( 65.10, 28.32){\circle*{0.75}}
 \put( 65.20, 28.30){\circle*{0.75}}
 \put( 65.30, 28.28){\circle*{0.75}}
 \put( 65.40, 28.26){\circle*{0.75}}
 \put( 65.50, 28.24){\circle*{0.75}}
 \put( 65.60, 28.23){\circle*{0.75}}
 \put( 65.70, 28.21){\circle*{0.75}}
 \put( 65.80, 28.19){\circle*{0.75}}
 \put( 65.90, 28.17){\circle*{0.75}}
 \put( 66.00, 28.16){\circle*{0.75}}
 \put( 66.10, 28.14){\circle*{0.75}}
 \put( 66.20, 28.12){\circle*{0.75}}
 \put( 66.30, 28.10){\circle*{0.75}}
 \put( 66.40, 28.08){\circle*{0.75}}
 \put( 66.50, 28.07){\circle*{0.75}}
 \put( 66.60, 28.05){\circle*{0.75}}
 \put( 66.70, 28.03){\circle*{0.75}}
 \put( 66.80, 28.01){\circle*{0.75}}
 \put( 66.90, 28.00){\circle*{0.75}}
 \put( 67.00, 27.98){\circle*{0.75}}
 \put( 67.10, 27.96){\circle*{0.75}}
 \put( 67.20, 27.95){\circle*{0.75}}
 \put( 67.30, 27.93){\circle*{0.75}}
 \put( 67.40, 27.91){\circle*{0.75}}
 \put( 67.50, 27.89){\circle*{0.75}}
 \put( 67.60, 27.88){\circle*{0.75}}
 \put( 67.70, 27.86){\circle*{0.75}}
 \put( 67.80, 27.84){\circle*{0.75}}
 \put( 67.90, 27.83){\circle*{0.75}}
 \put( 68.00, 27.81){\circle*{0.75}}
 \put( 68.10, 27.79){\circle*{0.75}}
 \put( 68.20, 27.78){\circle*{0.75}}
 \put( 68.30, 27.76){\circle*{0.75}}
 \put( 68.40, 27.74){\circle*{0.75}}
 \put( 68.50, 27.73){\circle*{0.75}}
 \put( 68.60, 27.71){\circle*{0.75}}
 \put( 68.70, 27.69){\circle*{0.75}}
 \put( 68.80, 27.68){\circle*{0.75}}
 \put( 68.90, 27.66){\circle*{0.75}}
 \put( 69.00, 27.64){\circle*{0.75}}
 \put( 69.10, 27.63){\circle*{0.75}}
 \put( 69.20, 27.61){\circle*{0.75}}
 \put( 69.30, 27.59){\circle*{0.75}}
 \put( 69.40, 27.58){\circle*{0.75}}
 \put( 69.50, 27.56){\circle*{0.75}}
 \put( 69.60, 27.55){\circle*{0.75}}
 \put( 69.70, 27.53){\circle*{0.75}}
 \put( 69.80, 27.51){\circle*{0.75}}
 \put( 69.90, 27.50){\circle*{0.75}}
 \put( 70.00, 27.48){\circle*{0.75}}

\end{picture}
\end{center}
\caption{The calculated dependence of $nS$ - charmonium leptonic
constants and the experimental values of $f_{\psi(nS)}$.}
\label{fh2}
\end{figure}

Second, eq.(\ref{2.3}) gives estimates of the leptonic constants for
the heavy $B$ and $D$ mesons, so these estimates are in a good agreement with
the values, obtained in the framework of other schemes of the QCD
sum rules \cite{3}.

These two facts show that the offered scheme can be quite reliably applied to
the systems with the heavy quarks.

Taking a value of the $1S$-level leptonic constant as the input one, we
calculate the leptonic constants of higher $nS$-excitations in the
charmonium and the bottomonium.

The results are presented in Tables \ref{th1}, \ref{th2} and on Figures
\ref{fh1}, \ref{fh2}. One can see that eq.(\ref{3}) is in a good agreement
with the experimental values of leptonic constants for the $nS$-wave
levels of heavy quarkonia \cite{9}.

One has to note, that for the $f_{\psi(3S)}$ value we have taken
\begin{equation}
f^2_{\psi(3S)} = f^2_{\psi(3770)}+f^2_{\psi(4040)}\;,
\end{equation}
since, as it generally accepted, the $\psi(3770)$ and $\psi(4040)$
states are the results of $3D$- and $3S$-levels mixing in the charmonium.

\section*{Conclusion}

In the framework of the QCD sum rules for the  leptonic constants
of the heavy quarkonia one uses the conditions of low $\Lambda_{QCD}/m_Q$
ratio value and the nonrelativistic quark motion in the phenomenological
potential, possessing the simple scaling properties, one takes into the account
the Coulomb-like $\alpha_S/v$ corrections, and in the scheme of the "smooth
average value" one derives the scaling expression, relating the leptonic
constants of $nS$-wave quarkonium levels, so
$$
\frac{f^2_{n_1}}{f^2_{n_2}} = \frac{n_2}{n_1}\;,
$$
independently of the heavy quark flavours.

The obtained relation is in a good agreement with the experimental
values of leptonic constants for the charmonium and the bottomonium, and
it reflects a small variation of the heavy quark kinetic energy
with respect to the heavy quark flavours.

\end{document}